\documentclass[twocolumn,pra,aps,superscriptaddress]{revtex4-2}
\usepackage[latin9]{inputenc}
\usepackage{mathtools}
\usepackage{amsbsy}
\usepackage{amssymb}
\usepackage{amstext}
\usepackage{bm}
\usepackage{xcolor}
\definecolor{darkblue}{rgb}{0, 0, 1}
\newcommand{\RN}[1]{
\bibliographystyle{unsrt}
	\textup{\uppercase\expandafter{\romannumeral#1}}
}
\usepackage[unicode=true,pdfusetitle,
bookmarks=false,
breaklinks=false,pdfborder={0 0 1},backref=false,colorlinks=true,allcolors=darkblue]
{hyperref}
\hypersetup{
	bookmarksnumbered=false,bookmarksopen=false}
\makeatletter
\@ifundefined{textcolor}{}
{
	\definecolor{BLACK}{gray}{0}
	\definecolor{WHITE}{gray}{1}
	\definecolor{RED}{rgb}{1,0,0}
	\definecolor{GREEN}{rgb}{0,1,0}
	\definecolor{BLUE}{rgb}{0,0,1}
	\definecolor{CYAN}{cmyk}{1,0,0,0}
	\definecolor{MAGENTA}{cmyk}{0,1,0,0}
	\definecolor{YELLOW}{cmyk}{0,0,1,0}
}

\newcommand{\beq}{\begin{equation}}
\newcommand{\eeq}{\end{equation}}
\newcommand{\beqa}{\begin{eqnarray}}
\newcommand{\eeqa}{\end{eqnarray}}

\usepackage{amsmath}
\usepackage{graphicx}
\usepackage{amssymb}
\usepackage{txfonts,color}
\makeatother

\begin{document}
\title{Weak-valued correlation functions: Insights and precise readout strategies}

\author{Yuan Feng}
\email{yuan.feng0630@gmail.com}
\affiliation{Blackett Laboratory, Imperial College London, Prince Consort Road, London, SW7 2AZ, United Kingdom}

\author{Xi Chen}
\email{chenxi1979cn@gmail.com}
\affiliation{Department of Physical Chemistry, University of the Basque Country UPV/EHU, Apartado 644, 48080 Bilbao, Spain}	
\affiliation{EHU Quantum Center, University of the Basque Country UPV/EHU, 48940 Leioa, Spain}

\author{Yongcheng Ding}
\email{jonzen.ding@gmail.com}
\affiliation{Department of Physical Chemistry, University of the Basque Country UPV/EHU, Apartado 644, 48080 Bilbao, Spain}

\date{\today}

\begin{abstract}	
	
{The correlation function in quantum systems plays a vital role in decoding their properties and gaining insights into physical phenomena. Its interpretation corresponds to the propagation of particle excitations between space-time, similar in spirit to the idea of quantum weak measurement in terms of recording the system information by interaction. By defining weak-valued correlation function, we propose the basic insights and the universal methods for recording them on the apparatus through weak measurement. To demonstrate the feasibility of our approach, we perform numerical experiments of perturbed quantum harmonic oscillators, addressing the intricate interplay between the coupling strength and the number of ensemble copies. Additionally, we extend our protocol to the domain of quantum field theory, where joint weak values encode crucial information about the correlation function. Hopefully, this comprehensive investigation can advance our understanding of the fundamental nature of the correlation function and weak measurement in quantum theories.}
\end{abstract}
\maketitle

\section{Introduction}
The correlation function is an indispensable aspect of quantum theory that furnishes a comprehensive characterization of the properties of quantum systems~\cite{GFJS}. By unifying essential  concepts, it serves as a powerful tool for investigating various quantities in quantum field theory, including scattering amplitudes, transition probabilities, and vertex amplitudes. Also, by employing Wick's trick, the correlation function provides insights into statistical behaviors in many-body physics~\cite{QSM}.  The physical interpretation for $n$-point correlator $\langle\Omega| \mathcal{T}[\hat{\phi}(\textbf{x}_1)\ldots\hat{\phi}(\textbf{x}_n)]|\Omega\rangle/\langle\Omega|\Omega\rangle$  corresponds to information extraction of the system with probing particles excitations. It is indeed the amplitude for a system preserving its ground state after (anti-)particle excitations at spacetimes $(\textbf{x}_1,...,\textbf{x}_n)$. Those excited particles could be viewed as probing particles or apparatuses due their 'weak' nature compared with the whole system. Besides the more conventional correlator, the out-of-time-correlator has also proven valuable in exploring engaging phenomena such as quantum chaos and information scrambling, which remain active frontiers of research~\cite{JHEPOTOC,JEOTOC,scrambling,OTOCtopo}. 

Meanwhile, recent developments in quantum information allows us to retrieve the information of a quantum system  through interacting with the apparatus. In particular, weak measurement of the Aharonov, Albert, and Vaidman (AAV) formalism~\cite{AAV} emerged as a novel way of recording the information on the pointer while weakly disturbing the system, sharing a similar idea with the interpretation of correlation function mentioned above.  Weak measurement lead to a range  of  cutting-edge experiment proposals~\cite{bomb,cheshire} and advancements~\cite{dondeestabas,foundation}, using this technique. These experiments include the record of physical quantities~\cite{geometric,eraser,discord}, quantum tomography~\cite{tomo,tomosc}, quantum steering~\cite{steering,gefen,sqc}, and transition detection~\cite{w2s,topological}.

This similarity of extracting information via weak interaction motivates us to consider the Weak-Valued Correlation Function, as the outcome of a weak measurement. Interestingly, this perspective is also supported by Gell-Mann and Low theorem, as well as Two-State Vector Formalism. In the calculation of correlation function, it is customary to invoke the Gell-Mann and Low theorem~\cite{GL} before embarking on Feynman diagram calculations. This theorem allows one to recast the $n$-point correlator $\langle\Omega| \mathcal{T}[\hat{\phi}(\textbf{x}_1)\ldots\hat{\phi}(\textbf{x}_n)]|\Omega\rangle/\langle\Omega|\Omega\rangle$ as a sum of connected diagrams with $n$ external lines.  The physical interpretation of it entails the evolution of two vacuum states $\langle 0||0\rangle$ from the infinite past and future to the state $\langle\Omega||\Omega\rangle$ in the present. If we adopt a time-symmetric perspective, two-state vector formalism (TSVF)~\cite{tsvf,tsvfreview} could be introduced as an alternative to the Gell-Mann and Low theorem. Without requiring adiabaticity, TSVF allows for arbitrary Hamiltonian to guide the evolution of eigenstates towards the pre-selection or post-selection state, offering us the weak-valued interpretation of the correlation function~\cite{sagawa}.

In this article, we define Weak-Valued Correlation Function using the standard AAV formalism, where the correlation function is viewed as the result of weak measurement. In Sec.~\ref{Weak Measurement as Excitations}, we propose the general ideas and universal schemes for recording the correlation function on the apparatus through weak measurement. In Sec.~\ref{PQHO}, we start with the perturbed quantum harmonic oscillator as the minimal example to exemplify our scheme.
By measuring eigenvalues, we estimate the final apparatus state, yielding complex readouts~\cite{jozsa}. Our findings reveal the trade-off between measurement strength and the number of experimental copies. This trade-off determines an optimal coupling coefficient, allowing the closest estimation of weak values with limited experimental copies. In Sec.~\ref{PHI4}, we further extend our analysis to quantum field theory utilizing the framework proposed by Dressel \textit {et~al.}~\cite{dressel}, which presents an alternative paradigm for implementing weak measurement. In Sec.~\ref{Discussion and Outlook}, we analyze the pivotal role of Gell-Mann and Low theorem and TSVF in weak-valued correlation function as part of motivation of our work. We also discuss the experimental realization and further investigation on correlators and quantum foundations.

\section{Weak-Valued Correlation Function}
\label{Weak Measurement as Excitations}

The basic idea of Weak-Valued Correlation Function (WVCF) is to construct the spacetime excitation of correlation function through weak measurement. This could be realized through measuring the excitation operator of the system defined by
\begin{align}\label{excitation operator}
\hat{G} = \mathcal{T}[\hat{\phi}(\textbf{x}_1)\ldots\hat{\phi}(\textbf{x}_n)]
\end{align}
under pre- and post- selection of the system into the ground state $|\Omega\rangle$.
The scheme of AAV-type weak measurement giving WVCF is as follows. We begin by preparing the system quantum state in the pre-selection state $|\Omega\rangle$, and the apparatus in an arbitrary state $|\phi\rangle$, which combined together as a separable state $|\Omega\rangle\otimes|\phi\rangle$.
Then, we couple the system and apparatus using an impulsive interaction
\begin{align}
\label{eq:Int}
H_{\text{int}}=g\delta(t-t_0)\hat{G}\otimes\hat{A},
\end{align}
where $\hat{A}$ is the Hermitian operator of the apparatus used to couple. Finally, after post-selection $|\Omega\rangle$ on the system, the whole state evolves into 
\begin{align}
\langle\Omega|\Omega\rangle \; \left[  | \Omega\rangle\otimes  e^{-igG_w\hat{A}} |\phi\rangle\right],
\end{align} 
in the weak coupling regime $g\rightarrow0$. $G_w$ here is usually called weak value for $\hat{G}$ and different spacetime $(\textbf{x}_1,...,\textbf{x}_n)$ gives different $\hat{G}$ thus different values of $G_w$. Thus, we define $G_w(\textbf{x}_1,...,\textbf{x}_n)$ as WVCF, which converges to the true value of correlation function in the limit of in the limit of $g\rightarrow0$:
\begin{align}
G_w(\textbf{x}_1,...,\textbf{x}_n) \rightarrow \frac{\langle\Omega| \mathcal{T}[\hat{\phi}(\textbf{x}_1)\ldots\hat{\phi}(\textbf{x}_n)]|\Omega\rangle}{\langle\Omega|\Omega\rangle},~g\rightarrow0.
\end{align}

To precisely readout the WVCF recorded in the final state of appratus, we resort to the average value of any operator $\hat{M}$ of the apparatus in the final state (derivation details are put in Appendix \ref{CWVD}):
\begin{align}\label{readout the WVCF}
\left\langle \hat{M}\right\rangle_f  & \simeq   \left\langle \hat{M}\right\rangle_i +
 ig\left\langle \hat{A}\hat{M}-\hat{M}\hat{A}\right\rangle _{i}\text{Re}G_w\notag
 \\&+g\left[\left\langle \hat{A}\hat{M}+\hat{M}\hat{A}\right\rangle _{i}-2\left\langle \hat{M}\right\rangle _{i}\left\langle \hat{A}\right\rangle _{i}\right]\text{Im}G_w,
\end{align}
where $\left\langle...\right\rangle_i$ corresponds to the average over initial state of the apparatus $|\phi\rangle$ and $\left\langle...\right\rangle_f$ corresponds to the average over final state of the apparatus $e^{-igG_w\hat{A}} |\phi\rangle$. Here Eq.~\eqref{readout the WVCF} corresponds to an equation set of two unknown variables $\text{Re}G_w$ and $\text{Im}G_w$, the solution of which gives us $G_w$.

Note that there are other well-known definitions of weak measurement, e.g., the continuous-in-time POVM~\cite{korotkov2001,jacobs2006}. However, the weak measurement in this paper only refers to the standard AAV formalism that consists of a pre-selection on the system, system-apparatus coupling, and a post-selection on the system, accumulating a weak value on the apparatus wave function.

Furthermore, our definition of WVCF and the weak measurement scheme could be directly generalised to quantum field systems by considering the system and apparatus as two separate degrees of freedom of a quantum field and by considering the pre- and post- selected states as the boundary conditions of the quantum field~\cite{dressel}. To define WVCF, we need to perform the polar decomposition $\hat{G}=\hat{U}\hat{R}$ (see section \ref{PHI4} for explanations), where $\hat{U}$ is a unitary operator and $\hat{R}$ is a Hermitian positive semidefinite operator~\cite{pati}. This time we measure the Hermitian part $\hat{R}$ of excitation operator:
\begin{align}
\label{eq:IntQFT}
H_{\text{int}}=g\delta(t-t_0)\hat{R}\otimes\hat{A},
\end{align}
under the pre-selection $|I\rangle=|\Omega\rangle\otimes|I_a\rangle$ and post-selection $|F\rangle=|\Theta\rangle\otimes| F_a\rangle$ where $|\Theta\rangle=\hat{U}^\dag|\Omega\rangle$.
According to the quantum action principle~\cite{QAP}, the variation of detection probability is given by:
\begin{align}
\label{eq:QAP}
\delta \ln \text{p}=2\delta g\left(\text{Re}R_w\text{Im}A_w+\text{Im}R_w\text{Re}A_w\right),
\end{align}
where $p=\left|\langle F|e^{-i\int \hat{H}_{\text{int}} dt}|I\rangle\right|^2$ is the detection probability of the quantum field and $A_w=\langle F_a|\hat{A}|I_a\rangle/\langle F_a|I_a\rangle$ is the weak value of $\hat{A}$. To extract weak value $R_w$, we perform measurements on the left side of Eq.~\eqref{eq:QAP}, utilizing suitable pre-selection $|I_a\rangle$ and post-selection $|F_a\rangle$ states for solving the corresponding linear equations. The solution of Eq.~\eqref{eq:QAP} gives us $R_w\rightarrow\frac{\langle\Theta|\hat{R}|\Omega\rangle}{\langle\Theta|\Omega\rangle}$ when $\delta g\rightarrow0$.  We call $R_w$ as weak-valued hermitian function (WVHF).  Accordingly, we define WVCF in quantum field as: 
\begin{align}\label{WVCF in QFT}
G_w=R_w\frac{\langle\Theta|\Omega\rangle}{\langle\Omega|\Omega\rangle}
\rightarrow\frac{\langle\Omega| \mathcal{T}[\hat{\phi}(\textbf{x}_1)\ldots\hat{\phi}(\textbf{x}_n)]|\Omega\rangle}{\langle\Omega|\Omega\rangle},~\delta g\rightarrow0.
\end{align}
The details of readout strategies in quantum field theory will be discussed in Sec.~\ref{PHI4}.

In this work, we focus on this type of measurement and refrain from discussing the detailed implementation on a specific quantum platform at this stage. Although the operator being measured may be a product of local operators, the response obtained through weak measurements aligns with the established theory of weak value~\cite{nonlocal}. Moreover, the theory remains valid even in cases where nonlocal interactions are necessary for certain tasks, enabling experimental devices to realize joint weak values~\cite{jwv}.

\section {Simplest Case: Perturbed Quantum Harmonic Oscillator}\label{PQHO}

Next, we exemplify our proposal in quantum mechanics, by considering the perturbed quantum harmonic oscillator (PQHO). 
The Hamiltonian reads $H=p^2/2m+m\omega^2 x^2/2+\lambda x^{4}$, where the anharmonic perturbation $\lambda x^{4}$ is treated as the self-interaction term. 

Without loss of generality, we define the excitation operator at an arbitrary time $t\geq 0$:
\begin{align}
 \hat{G}=x(t)x(0)=e^{iHt}xe^{-iHt}x,
 \end{align}
which its weak values correspond to the correlation function of PQHO.

To perform the weak measurement, we employ a qubit as the apparatus to record the weak value. We prepare the separable initial state
$
|\Omega\rangle\otimes\left[\cos\left(\frac{\theta_0}{2}\right) |\uparrow\rangle  +  \sin\left(\frac{\theta_0}{2}\right) e^{i\varphi}|\downarrow\rangle\right],
$
where $|\uparrow\rangle$ and $|\downarrow\rangle$ are the eigenbasis of $\hat{\sigma_z}$. Then, we couple the excitation operator of the system and the spin-y operator of the apparatus using an impulsive interaction
$H_{\text{int}}=g\delta(t-t_0)\hat{G}\otimes\hat{\sigma}_{y}$,
which entangles the quantum states of the system and apparatus. After the post-selection $|\Omega\rangle$, the WVCF accumulates on the final state of the apparatus, given by $e^{-igG_w\hat{\sigma}_{y}} |\phi\rangle$.
It is worth noting that in the limit of $g\rightarrow0$, they converge to the true values.

The physical interpretation of the WVCF  corresponds to a rotation through a small angle $(\theta-\theta_0)=2gG_w$ around the $Y-$axis, assuming $G_w$ is real. 
In the case of a complex weak value $G_w=\text{Re}G_w+i\text{Im}G_w$~\cite{jozsa}, we need to analyze the effect of the non-unitary operator $e^{-igG_w\hat{\sigma}_{y}}$ on the apparatus, to retrieve the value of correlator at arbitrary time. In the weak coupling regime, the expectation values of the qubit along different directions satisfy the following equations (see Appendix \ref{CWVD} for details)
\begin{align}
\label{eqn:complex}
\langle\hat{\sigma}_{y}\rangle _{f}& \simeq\langle \hat{\sigma}_{y}\rangle _{i}+2g\left[1-\langle \hat{\sigma}_{y}\rangle _{i}^{2}\right]\text{Im}G_w,
\notag\\
\langle \hat{\sigma}_{x}\rangle _{f}& \simeq\langle \hat{\sigma}_{x}\rangle _{i}+2g\langle \hat{\sigma}_{z}\rangle _{i}\text{Re}G_w-2g\langle \hat{\sigma}_{x}\rangle _{i}\langle \hat{\sigma}_{y}\rangle _{i}\text{Im}G_w,
\notag\\
\langle \hat{\sigma}_{z}\rangle _{f}& \simeq\langle \hat{\sigma}_{z}\rangle _{i}-2g\langle \hat{\sigma}_{x}\rangle _{i}\text{Re}G_w-2g\langle \hat{\sigma}_{z}\rangle _{i}\langle \hat{\sigma}_{y}\rangle _{i}\text{Im}G_w.
\end{align}
By measuring the expectation values of the spin components, we can access the value of $G_w$. With the estimations of $\langle\hat{\sigma}_y\rangle_f$ and $\langle\hat{\sigma}_x\rangle_f$, we estimate the WVCF as follows
\begin{align}
\label{ab}
\text{Im}G_w &= \frac{\langle\hat{\sigma}_y\rangle_f-\langle\hat{\sigma}_y\rangle_i}{2g\left[1-\langle\hat{\sigma}_y\rangle_i^2\right]},
\notag\\
\text{Re}G_w &= \frac{\langle\hat{\sigma}_x\rangle_i\langle\hat{\sigma}_y\rangle_i}{\langle\hat{\sigma}_z\rangle_i} \text{Im}G_w + \frac{\langle\hat{\sigma}_x\rangle_f-\langle\hat{\sigma}_x\rangle_i}{2g\langle\hat{\sigma}_z\rangle_i}.
\end{align}

 \begin{figure}
\includegraphics[width=8.6cm]{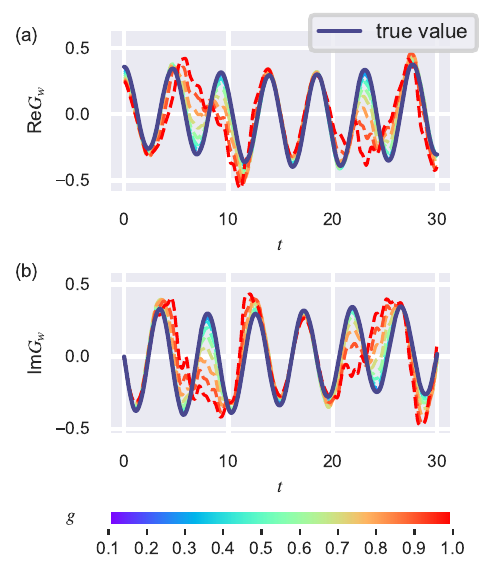}
\caption{\label{fig:acc_g} Weak-valued correlation function determined from Eq.~\eqref{ab} for different measurement strength $g$ during the time $t\in[0,30)$, plotted in (a) real part and (b) imaginary part, respectively. The true values of the correlation function are represented by a dark solid line, while the weak values are depicted as a rainbow series of dotted lines.}
\end{figure}

Fig.~\ref{fig:acc_g} provides valuable insight into the behavior of our protocol in the PQHO model, highlighting the importance of operating within the weak coupling regime to ensure accurate readout of the WVCF. The readout of our protocol is calculated through numerical simulations for different coupling strengths $g$ according to Eq.~\eqref{ab}. Here, we approximate the perturbed ground state as  $|\Omega\rangle\simeq|0\rangle+[(-3/2\sqrt{2})|2\rangle-(\sqrt{6}/8)|4\rangle]\lambda$ in the eigenbasis of the harmonic oscillator $|n\rangle$ and we express the $\hat{G}$ in the basis of $|\Omega\rangle$ as a 6-dimensional operator (see Appendix \ref{GSGL} and \ref{TEHO} for details). 
We consider the calculated results under 6-dimensional cutoff as the true values.

By observing the expectation values $\langle\hat{\sigma}_x\rangle_f$ and $\langle\hat{\sigma}_y\rangle_f$ and solving Eqs.~\eqref{eqn:complex}, which hold only in the weak coupling regime, we retrieve the WVCF. Our simulations demonstrate that the readout deviates significantly from the true value at larger value of $g$, indicating the occurrence of the weak-to-strong transition \cite{w2s}. In this regime, the higher-order terms in the expansion of the evolving operator $\exp(-ig\hat{G}\otimes\hat{\sigma}_y)$ introduce non-negligible effects on the apparatus. However, we establish the validity of our protocol by showing that the readouts at $g=0.1$ exhibit good agreement with the true values, confirming the accuracy of our approach. 
We underscore the significance of adhering to the weak coupling regime for reliable and precise readout of the WVCF.

\begin{figure*}
\includegraphics[width=17.2cm]{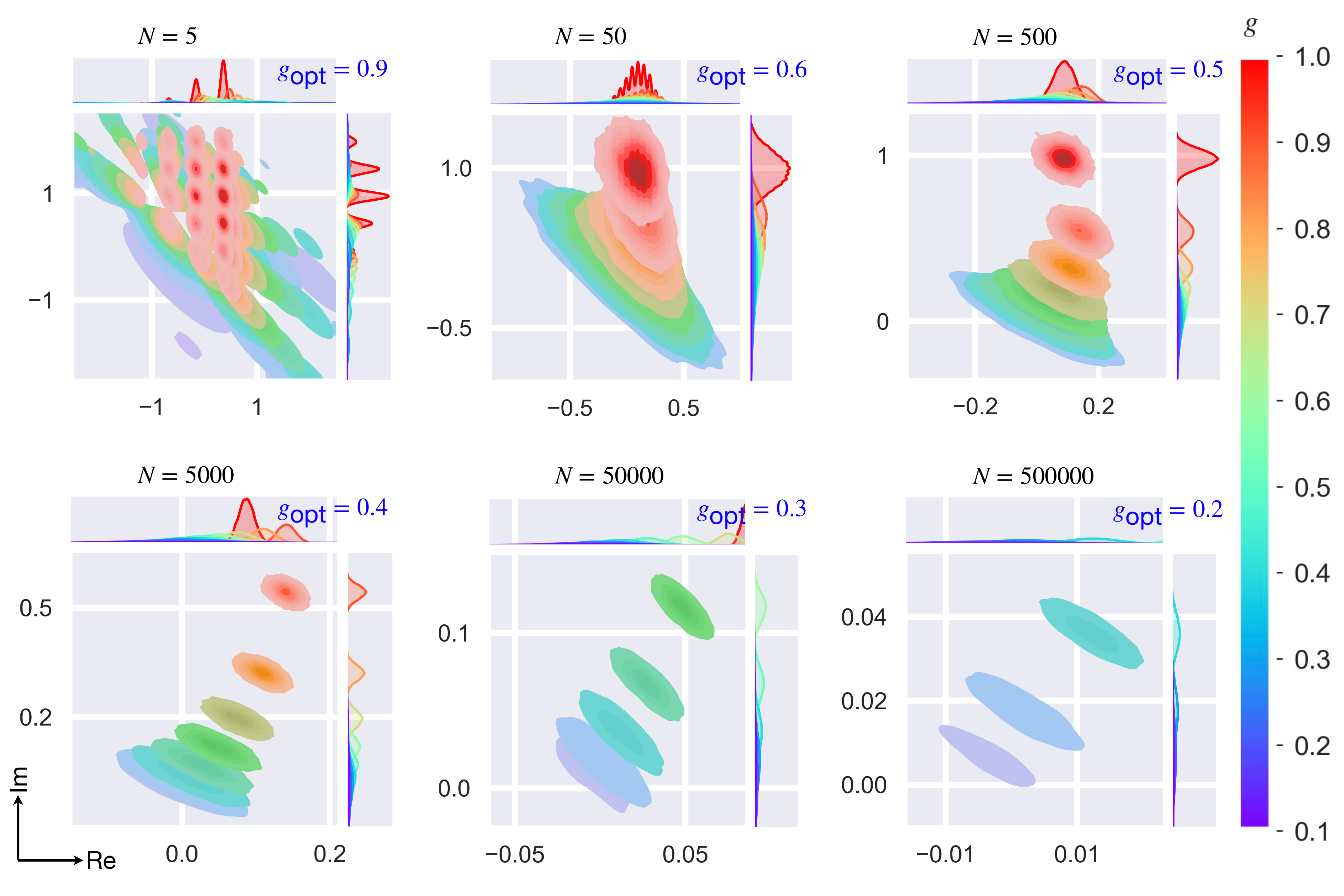}
\caption{\label{fig:tradeoff}
Gaussian kernel density distribution of relative radical vector $z_{\text{RV}}=\left(z_{\text{R}}-z_{\text{TV}} \right) / |z_{\text{TV}}|$ is presented on the complex plane, for various numbers of copies $2N$ and coupling strength $g$. The marginal density distributions are shown on the the up and right sides, providing additional information about the distribution characteristics. In the upper right corner of each subfigure, we indicate the optimal value of the coupling strength $g$ for each corresponding $N$.}
\end{figure*}

It is important to note that the simulations presented in Fig.~\ref{fig:acc_g} are idealized, assuming perfect estimations of the expectation values $\langle\hat{\sigma_i}\rangle_f$, i.e.,  requiring infinite copies of the final state $e^{-igG_w\hat{\sigma}_{y}} |\phi\rangle$. In experimental implementations, however, the accuracy of our readout depends on both the coupling strength and the number of available copies. Let us consider that we have access to $2N$ copies of the final state in a single experiment, allowing us to perform $N$ shots of measurement on the operators $\hat{\sigma}_x$ and $\hat{\sigma}_y$, respectively. By averaging the eigenvalues $ \sigma_i =\pm1$ as the measurement outcomes, we obtain the imperfect estimations of $\langle\hat{\sigma_i}\rangle_f$ as $\langle\hat{\sigma_i}\rangle_E=\sum_{i=1}^N \sigma_i $. The discrepancy between $\langle\hat{\sigma_i}\rangle_f$ and $\langle\hat{\sigma_i}\rangle_E$ is bounded by $|\langle\hat{\sigma_i}\rangle_f-\langle\hat{\sigma_i}\rangle_E|\leq\varepsilon(N,\delta)$, where $\varepsilon(N,\delta)$ represents the upper bound on the measurement-induced deviation with a probability of $1-\delta$. Several explicit bounds can be used, such as the empirical Bernstein bound (EBB)~\cite{bernstein}, which is applicable in various cases. Moreover, in practical scenarios, the optimal coupling strength is not necessarily the weakest possible. When the signal-to-noise ratio decreases, extracting meaningful information from the apparatus becomes more challenging, leading to less precise retrieval of the correlator value with a limited number of copies $N$. Conversely, a stronger coupling enables a more accurate estimations of $\langle\hat{\sigma_i}\rangle_f$, albeit with a larger deviation between the weak value and the true value. Thus, a trade-off exists between the coupling strength $g$ and the number of copies $2N$. Depending on specific criteria, one can select an appropriate coupling strength for a given number of copies, aiming to achieve a readout that sufficiently approaches the true values.

\begin{figure}
\includegraphics[width=8.6cm]{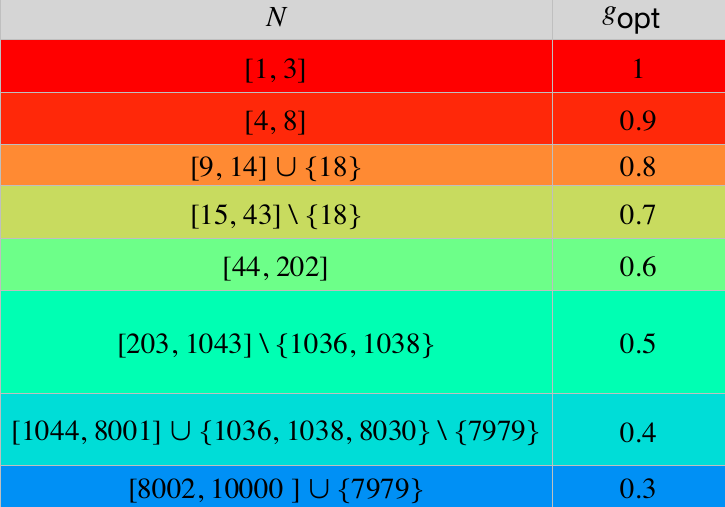}
\caption{\label{Detailed trade-off data} Detailed trade-off data of optimal $g\in G=\left\{0.1, 0.2, ..., 1\right\}$ for each $N\in\left[1,10000\right]$.}
\end{figure}

 \begin{figure*}
\includegraphics[width=17.2cm]{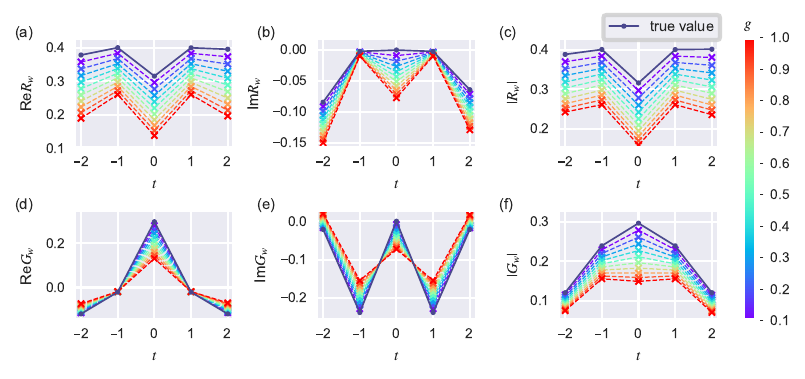}
\caption{\label{fig:qft}  Numerical simulations of a $\phi^4$ lattice field theory. On the upper panels are the weak-valued effective hermitian function $R_w$ of a $\phi^4$ lattice field theory determined from Eq.~\eqref{WMRw} for different measurement strength $g$ during the time  $t\in\{-2,-1,...,2\}$, plotted in (a) real part, (b) imaginary part, and (c) module. On the lower panels are the corresponding weak-valued correlation function obtained from $G(t,0)=R_w\langle\Theta|\Omega\rangle / \langle\Omega|\Omega\rangle$, plotted in (d) real part, (e) imaginary part, and (f) module. We plot the true values of the correlation function as dark dots, while weak values as crosses colored in rainbow. }
\end{figure*}

Following this spirit, we design a numerical experiment to retrieve the WVCF  in the practical situation. The main results of this experiment simulation are shown in Fig.~\ref{fig:tradeoff},  where we analyze  the distribution of the relative error radical vector $z_{\text{RV}}=\left(z_{\text{R}}-z_{\text{TV}} \right) / |z_{\text{TV}}|$ on the complex plane. Here, $z_{\text{R}}$ is the readout of weak value based on the estimations $\langle\hat{\sigma_i}\rangle_E$ and $z_{\text{TV}}$ is the true value of the correlator. We vary the coupling strength from weak $(g=0.1)$ to strong $(g=1)$ with an interval of $0.1$, and examine the variation trend in distribution of the radical vector $z_{\text{RV}}$ by increasing the number of copies $2N$.  It is  observed that with increasing $2N$, the distribution changes from a multiple-peak-like shape to a Gaussian distribution on the complex plane. In the few-shot cases ($N=5, 50$), the discrete nature of the values of possible readouts, resulting from the estimations $\langle\hat{\sigma_i}\rangle_E$, is blurred by performing Gaussian interpolation before plotting them as dots for illustrative purposes. As $N$ increases significantly, the readout values tend to converge to a continuum limit, approximating a complex normal distribution. To ensure reliable distributions, we perform $M=10000$ individual simulations with specific settings ${g_i, N_j}$ to minimize statistical fluctuations. In terms of the trade-off, our objective is to identify the coupling strength that maximizes the likelihood of minimal deviation between the readout and the true value in a single experiment. After analyzing the raw data, we indicate the optimal coupling strength for each $N\in [1,10000]$ in Fig~\ref{Detailed trade-off data}. (see methodology and additional information in Appendix~\ref{NSDTD}). This numerical experiment yields valuable insights into the behavior of our protocol, aiding in the determination of the appropriate coupling strength and number of copies for accurately reading the WVCF.

\section {Extension to QFT: $\phi^4$ theory}\label{PHI4}

In the regime of weak interaction, the correlation function could be regarded as the weak value of field operators products under appropriate pre- and post-selected boundary conditions~\cite{dressel}. We hereby extend the concept of WVCF to quantum field theory by considering the interaction between two separate degrees of freedom of the field under the pre- and post- spacetime boundary conditions. We take $\phi^4$ as an example to illustrate this extension. The Lagrangian density of the  $\phi^4$ theory is written by  $\mathcal{L}=\frac{1}{2}\partial^\mu\phi\partial_\mu\phi-\frac{1}{2}m^2\phi^2-\frac{\lambda}{4!}\phi^4$.

We consider the excitation operator at the origin and its polar decomposition:
\begin{align}
\hat{G}=\mathcal{T} \phi(t,0) \phi(0,0)=\hat{U}\hat{R}.
\end{align}
Since the quantum action principle Eq.~\eqref{eq:QAP} only allows Hermitian operators, we need to separate the Hermitian part of the excitation operator using polar decomposition $\hat{G}=\hat{U}\hat{R}$~\cite{pati}. We denote the ground state $|\Omega\rangle$ and after unitary transformation $|\Theta\rangle=\hat{U}^{\dagger}|\Omega\rangle$.  We consider the orbital degree of freedom of the local field as the system and the spin degree of freedom of the local field as the apparatus, which are two separate degrees of freedom of a local field. To carry out the weak measurement, we couple the Hermitian part of the excitation operator and the Pauli-x operator of the field  $\hat{H}_{\text{int}}=g\delta(t-t_0)\hat{R}\otimes\hat{\sigma}_x$ under the initial  $|I\rangle=|\Omega\rangle\otimes|I_a\rangle$ and final states of the field $|F\rangle=|\Theta\rangle\otimes| F_a\rangle$. By applying the the quantum action principle~\cite{GFJS,QAP}, we have the joint weak value of the the Hermitian variation of the quantum action $\delta\hat{S}=-\delta g\hat{R}\otimes\hat{\sigma}_x$, which encodes the amplitude $a=\langle F|e^{-i\int \hat{H} _{\text{int}} dt}|I\rangle=\langle\Theta|\otimes\langle F_a|e^{-ig\hat{R}\otimes\hat{\sigma}_x}|\Omega\rangle\otimes|I_a\rangle$ as a weak value $S_w=\langle F| \delta\hat{S}|I\rangle/\langle F|I\rangle=-i\hbar\delta\ln a$, giving Eq.~\eqref{eq:QAP}. To get the $R_w$, we choose $|I_a\rangle_1=|\uparrow\rangle $, $|F_a\rangle_1=1/\sqrt{2} \left(|\uparrow\rangle+|\downarrow \rangle \right)$ and $|I_a\rangle_2=1/\sqrt{2} \left(i|\uparrow\rangle+|\downarrow\rangle \right) $, ${ |F_a\rangle_2= |\downarrow\rangle} $ for convenience,
\begin{align}\label{WMRw}
\text{Re}\left\{ R_{w}\right\} &= \delta\ln \text{p}_{2}/{2\delta g}\notag,\\
\text{Im}\left\{ R_{w}\right\} &= \delta\ln \text{p}_{1}/{2\delta g}.
\end{align}
where $\text{p}=|a|^2$ is the detection probability and $\delta g$ is the tunable coupling strength. We can retrieve the WVCF $G_w=R_w{\langle \Theta|\Omega\rangle}/{\langle\Omega|\Omega\rangle}$ by multiplying $R_w$ with ${\langle \Theta|\Omega\rangle}/{\langle\Omega|\Omega\rangle}$, which interestingly coincides with the amplitude between the selection on the first degree of freedom of the local field.

To demonstrate the protocol, we simulate the weak-valued two-time correlation function of a $\phi^4$ lattice field theory with $(1+1)$-dimensional spacetime in Fig.~\ref{fig:qft}. 
Based on the quantum action principle, it provides an alternative paradigm for weak measurement, differing from the standard AAV formalism. As an infinitesimal variation, this paradigm is closely related to the path integral formalism and can be applied in quantum mechanics. Refer to Appendix~\ref{LSPF} for all simulation details.

\section {Discussion and Outlook}\label{Discussion and Outlook}
 
After formulating WVCF and its readout strategies in quantum mechanics and quantum field theory, we hereby present the discussion on the equivalence between TSVF and GL theorem in the interpretation of correlation functions. It also serves as the inspiration of this work. Following the GL theorem, we derive ground states $|\Omega^{\mp} \rangle = \lim_{\epsilon\rightarrow 0} U_{\epsilon I}(0,\mp\infty)|0\rangle/\langle0|U_{\epsilon I }(0,\mp \infty)|0\rangle$ by adiabatically evolving the vacuum state $|0\rangle$ from infinite past/future to present. Here $U_{\epsilon I}(0,\mp\infty)$ is the time evolution operator corresponding to Hamiltonian $H_\epsilon(t)=H_0 +e^{-\epsilon |t|}\lambda V$, which is known as M\o ller operator~\cite{Mollar}. Thus, we write down the $n$-point correlation function of the corresponding system $G^{(n)}(\textbf{x}_1,\cdots,\textbf{x}_n)= \langle 0|\mathcal{T}[\hat{\phi}_I(\textbf{x}_1)...\hat{\phi}_I(\textbf{x}_n)\mathcal{S}]|0\rangle /\langle 0|\mathcal{S}|0\rangle=\langle\Omega^+| \mathcal{T}[\hat{\phi}(\textbf{x}_1)\ldots\hat{\phi}(\textbf{x}_n)]|\Omega^-\rangle/\langle\Omega^+|\Omega^-\rangle$, satisfying the adiabatic criteria to preserve the ground state. Note that GL theorem is not compulsory to derive the same expression. For example, we have the an alternative way to arrive the ground state from the vacuum state, departing from finite past $t_1$ and future $t_2$ to the pre-selected state $|\Omega^-\rangle=U(t_{\text{now}},t_1)|0\rangle$ and post-selected state $\langle\Omega^+|= \langle 0|U^\dag(t_2,t_{\text{now}})$ at a certain time $t_{\text{now}}$. Accordingly, one can introduce an auxiliary potential in the total Hamiltonian to compensate for excitations or design a schedule for turning on the perturbation. Propagators $U(t_{\text{now}},t_1)$ and $U^\dag(t_2,t_{\text{now}})$ serve as shortcuts-to-adiabaticity~\cite{rev,adolfo} since it provides a more efficient way than the adiabatic evolution in GL theorem. Thus, one derives the same expression of correlation function, which can be understood as a weak value. 

Considering the experimental implementation, one may turn to digital quantum simulation, a flexible method for exploring quantum dynamics of both quantum mechanics and quantum field theory~\cite{dirac,adscft}. This approach allows entanglement and selections with quantum circuits. Alternatively, one also has the standard technique in quantum field theory by introducing weak source currents linearly coupled to the field in the Lagrangian, perturbing the field evolution within local apparatuses. The joint averaged response of these apparatuses reveals the desired correlators, shedding light on the dynamics and properties of the investigated quantum field system.

Going beyond the examples in the work, a compelling application lies in reading out OTOCs using weak measurement, where operators are non-time-ordered. Additionally, we can explore scattering matrix modulation with different pre- and post-selections to study non-equilibrium physics and quantum nonlocality through propagators. Meanwhile, it is also compatible to Keldysh formalism~\cite{keldysh} that extends backward propagation in the calculation of generalized correlation function, providing a systematic framework for investigating non-equilibrium systems~\cite{noneqprb2003,noneqprd2014,noneqprb2016}.

\section {Conclusion}
In summary, we introduce WVCF in quantum system and quantum field as the result of weak measurement. We propose a standard weak measurement of the AAV type to read out the WVCF of the PQHO. Our results reveal the interplay between the coupling strength and the number of copies in the readout. Furthermore, we extend our framework to quantum field theory, where correlator calculations are crucial. Finally, we redefine the Gell-Mann and Low theorem to develop a method for calculating correlation function. By employing TSVF to accelerate adiabatic evolution, we establish an equivalent theory where correlation function is interpreted as weak values.

\begin{acknowledgements}
Discussions with Justin Dressel, Yiming Pan, and Jie Lu are appreciated. This work is supported by the Basque Government through Grant No. IT1470-22, the Project Grants No. PID2021-126273NB-I00, No. PID2021-123131NA-I00 funded by MCIN/AEI/10.13039/501100011033 and by "ERDF A way of making Europe" and "ERDF Invest in your Future". This project has also received support from the Spanish Ministry of Economic Affairs and Digital Transformation through the QUANTUM ENIA project call-Quantum Spain, and by the EU through the Recovery, Transformation and Resilience PlanNextGenerationEU within the framework of the Digital Spain 2026 Agenda. X.C. acknowledges ayudas para contratos Ram\'{o}n y Cajal-2015-2020 (RYC-2017-22482).
\end{acknowledgements}

\appendix
\section{Ground state from Gellmann-Low theorem}\label{GSGL}

To calculate the ground state $|\Omega\rangle$ of the perturbed PQHO using the Gell-Mann and Low theorem, we consider the adiabatic Hamiltonian given by:
\begin{align}
 H_\epsilon(t)=\frac{p^{2}}{2m}+\frac{1}{2}m\omega^2 x^{2}+\lambda  e^{-\epsilon |t|} x^{4}.
\end{align}
We begin by focusing on the ground state $|\Omega^-\rangle$. Expanding the state in terms of a perturbation series, we express the target state as the ratio of two sets of infinite series:
\begin{align}\label{Omega from G-L}
	|\Omega^-\rangle = \lim_{\epsilon\rightarrow 0} \frac{U_{\epsilon I}(0,-\infty)|0\rangle}{\langle0|U_{\epsilon I }(0, - \infty)|0\rangle} = \lim_{\epsilon\rightarrow 0} \frac{\sum_{n=0}^{\infty}c_n \lambda^n}{\sum_{n=0}^{\infty}b_n \lambda^n} = \sum_{n=0}^{\infty} a_n \lambda^n,
\end{align}
where $U_{\epsilon I}$ is the adiabatic time evolution operator in the interacting picture. The series division on the right-hand side is implemented as:
$$
a_n = {1}/{b_0} \left[ c_n - a_0 b_n -...- a_{n-1} b_1\right],~\left( n=0,1,2...\right).
$$
Here, $a_n$ represents the coefficients of the perturbation series expansion of the ground state, which we aim to determine. The calculation involves solving a recursive equation to obtain the values of $a_n$. By following this procedure, we can determine the ground state of the PQHO Hamiltonian using the Gell-Mann and Low theorem.

Indeed, complex analysis reveals that the series representation $|\Omega^{-}\rangle = \sum_{n=0}^{\infty} a_n^{(\epsilon)} \lambda^n$ is a divergent series, with an asymptotic expansion only valid to the first order. On the complex plane, considering $|\Omega^{-}\rangle$ as a complex function of $\lambda$, it is non-analytic everywhere except at the origin. This can be understood by considering the case where $\lambda<0$. In this scenario, there is no minimum energy for the potential, resulting in the non-existence of the ground state $|\Omega^{-}\rangle$. Consequently, there is no circle of convergence outside the origin point.

\begin{figure} 
\includegraphics[width=8.6cm]{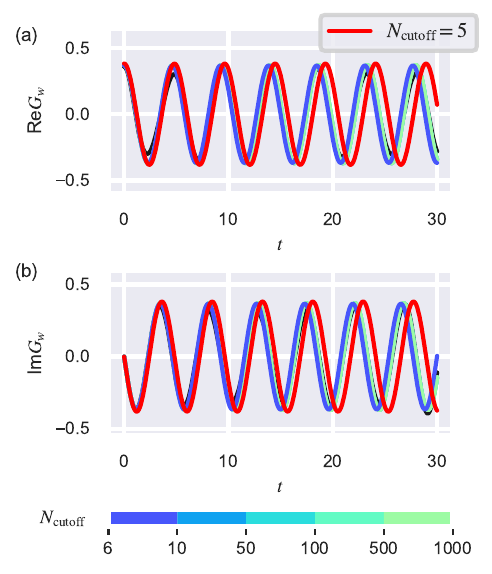}
\caption{\label{S1}  Correlation function simulated under different cutoff on energy level $N_{\text{cutoff}}$. The rainbow curves are used to show numerical simulation of solving the eigenequations of the Hamiltonians under the $N_{\text{cutoff}}$ shown in the subfigure above. The black curve shows the result calculated from Eq.~\eqref{ground state result} under a certain large enough cutoff energy level $N_{\text{cutoff}}=1000$, which is almost covered by the rainbow curves at large $N_{\text{cutoff}}$. Here we utilize complementary colors to explicitly enhance the distinction between $N_{\text{cutoff}}=5$ and $N_{\text{cutoff}}=6$. 
 }
\end{figure}

The order of asymptotic expansion is determined in the following way. In terms of Dyson expansion, time evolution operator is written as:
\begin{align*}
U_{\epsilon I}&(t,t_{0})=1+(-i)\int_{t_{0}}^{t}dt_{1}H_{\epsilon I}(t_{1})
\\
&+(-i)^{2}\int_{t_{0}}^{t}dt_{1}\int_{t_{0}}^{t_{1}}dt_{2}H_{\epsilon I}(t_{1})H_{\epsilon I}(t_{2})+...,
\end{align*} 
where $H_{\epsilon I}(t)=\lambda e^{iH_{0}t}e^{-\epsilon\left|t\right|}x^{4}e^{-iH_{0}t}$ is the the adiabatic Hamiltonian in the interaction picture. Acting on the $|0\rangle$ and inserting identity operator $1=\sum_n |n\rangle \langle n|$, it leads to:
\begin{eqnarray*}
\label{time evolution operator}
U_{\epsilon I}(t,t_{0})~|0\rangle &=& |0\rangle  + (-i)\lambda \sum_n \frac{\langle n |x^4|0\rangle}{in+\epsilon}|n\rangle \nonumber\\&+&(-i)^2 \lambda^2 \sum_{n,m} \frac{\langle n |x^4|m\rangle\langle m |x^4|0\rangle}{\left(in+2\epsilon \right) \left(im+\epsilon \right)}|n\rangle+...,
\end{eqnarray*} 
where $|n\rangle$ is the eigenbasis of harmonic oscillator satisfying $\hat{H_0}|n\rangle=E_n |0\rangle$ and $E_n = n+\frac{1}{2}$. 
The ground state could be derived by putting this into Eq.~\eqref{Omega from G-L} and calculating the series division:
\begin{align}
\label{ground state result}
\mid\Omega^-\rangle &=\sum_{n}a_{n}\lambda^{n}=a_{0}+a_{1}\lambda+a_{2}\lambda^{2}+...,
\notag\\
a_{0} &=\langle x\mid0\rangle,~
\notag\\a_{1} &= -\frac{3}{2\sqrt{2}} |2\rangle -\frac{\sqrt{6}}{8}  |4\rangle ,~
\notag\\a_{2}&= \frac{75}{8\sqrt{2}}|2\rangle + \frac{9\sqrt{6}}{4}|4\rangle + \frac{17}{16}\sqrt{5} |6\rangle + \frac{3\sqrt{70}}{64}|8\rangle.
\end{align} 
To find the optimal order for asymptotic expansion, we study correlation function at $t=0$: $G^{(2)}(0,0)={\langle\Omega|x^2|\Omega\rangle}/{\langle\Omega|\Omega\rangle}$. Our simulation result of solving the eigenequations gives $0.3700$. Reserving Eq.~\eqref{ground state result} to 1-order gives 0.3570, which is close to simulation result. However, when reserved to 2-order, it becomes 0.8484, with more deviation. As expected, the higher order it is, the greater deviation we get. Likewise, we get $|\Omega^+\rangle$  in a similar way $|\Omega^+\rangle=|\Omega^-\rangle=|\Omega\rangle$. As a result, 1-order is the optimal asymptotic series for $|\Omega\rangle$:
\begin{align}
|\Omega\rangle \simeq |0\rangle+\left[-\frac{3}{2\sqrt{2}}|2\rangle-\frac{\sqrt{6}}{8}|4\rangle\right]\lambda.
\end{align}

\section{Truncated eigenbasis of harmonic oscillator}\label{TEHO}

\begin{figure*}
\includegraphics[width=17.2cm]{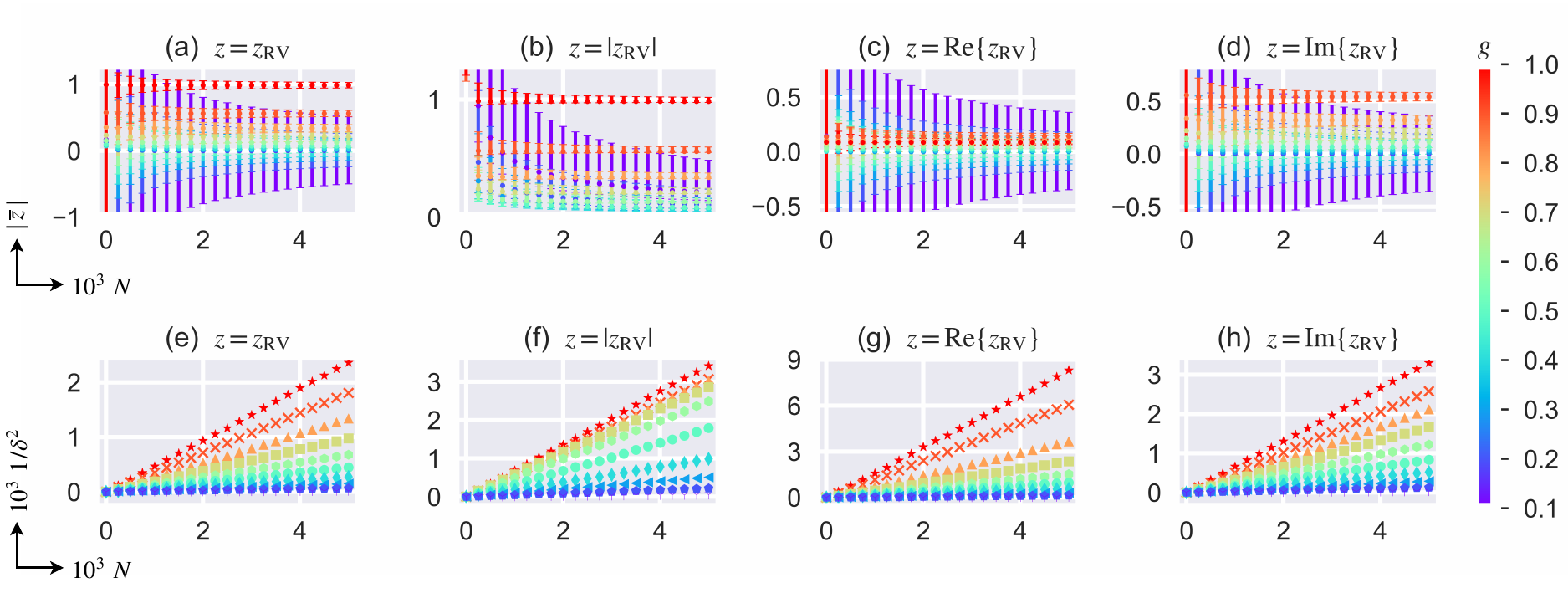}
\caption{\label{figS2} On the upper panels are the absolute mean values $|\overline{z}|$ for $M=10000$ sampling points of (a) complex value $z_{\text{RV}}=\left(z_{\text{R}}-z_{\text{TV}} \right) / |z_{\text{TV}}|$, (b) module $|z_{\text{RV}}| =|z_{\text{R}}-z_{\text{TV}}| / |z_{\text{TV}}|$, (c) real part $\text{Re}\left\{z_{\text{RV}}\right\} =\text{Re}\left\{z_{\text{R}}-z_{\text{TV}}\right\} / |z_{\text{TV}}|$ and (d) imaginary part $\text{Im}\left\{z_{\text{RV}}\right\} =\text{Im}\left\{z_{\text{R}}-z_{\text{TV}}\right\} / |z_{\text{TV}}|$. Error bars denote confidence intervals of 0.95. On the lower panels are uncertainty versus the number of copies: $1/\delta^2-N$ for $g\in G=\left\{0.1, 0.2, ..., 1\right\}$, over $M=10000$ sampling points of (e) complex value $z_{\text{RV}} =\left(z_{\text{R}}-z_{\text{TV}} \right) / |z_{\text{TV}}|$, (f) module $|z_{\text{RV}}| =|z_{\text{R}}-z_{\text{TV}}| / |z_{\text{TV}}|$, (g) real part $\text{Re}\left\{z_{\text{RV}}\right\} =\text{Re}\left\{z_{\text{R}}-z_{\text{TV}}\right\} / |z_{\text{TV}}|$ and (h) imaginary part $\text{Im}\left\{z_{\text{RV}}\right\} =\text{Im}\left\{z_{\text{R}}-z_{\text{TV}}\right\} / |z_{\text{TV}}|$. The relation $1/\delta^2-N$  proves to be linear in all the four cases.}
\end{figure*}

In order to address the challenge of infinite dimensionality in the calculation of the operator $\hat{A}=x(t)x(0)=e^{i\hat{H}t}xe^{-i\hat{H}t}x$, a truncation approach is employed. This approach involves selecting a cutoff energy level $N_{\text{cutoff}}$ that ranges from a sufficiently large value down to zero. For the sake of better illustration, we simulate the eigenequations of the Hamiltonians under the sufficient large cutoff on energy level $N_{\text{cutoff}}=1000$.

By expressing the operator $\hat{A}$ in terms of the eigenbasis of the harmonic oscillator, the cutoff energy level determines the range of energy levels considered in the calculation.  As shown in Fig.~\ref{S1}, the behavior of the correlation function $G^{(2)}(t,0)$ is observed as the cutoff energy level decreases.  It is noted that as $N_{\text{cutoff}}$ is reduced to $N_{\text{cutoff}}=6$, the result from our calculation Eq.~\eqref{ground state result} matches with the simulation result and the value of $G^{(2)}(0,0)$ remains constant at $0.35697259$. However, a turning point is reached when $N_{\text{cutoff}}$ reaches $N_{\text{cutoff}}=5$, resulting in  obvious deviation from simulation result and $G^{(2)}(0,0)=0.34803337$. This observation can be understood by considering that the calculated ground state is only relevant to the first five energy levels.

Based on these observations, a decision is made to choose $N_{\text{cutoff}}=6$ as the truncated energy level for the calculation. This choice ensures that the relevant energy levels are included while controlling the computational complexity associated with the infinite-dimensional space.

\section{Complex weak value through the spin of the pointer}\label{CWVD}

For any complex weak value $G_{w}=a+bi$ measured through interaction $\hat{G}\otimes \hat{A}$, any observable $\hat{M}$ of apparatus satisfies:
\begin{align}
\label{cwv-m}
\left\langle \hat{M}\right\rangle _{f}=\left\langle \hat{M}\right\rangle _{i} &+iga\left\langle \hat{A}\hat{M}-\hat{M}\hat{A}\right\rangle _{i}\notag
\\&+gb\left[\left\langle \hat{A}\hat{M}+\hat{M}\hat{A}\right\rangle _{i}-2\left\langle \hat{M}\right\rangle _{i}\left\langle \hat{A}\right\rangle _{i}\right],
\end{align}
where we assume $\hat{A}$ is Hermitian. This equation is derived following the method in~\cite{jozsa}. The initial and final state of apparatus are $|\phi\rangle$ and $|\alpha\rangle = e^{-igG_w\hat{A}} |\phi\rangle \simeq \left(1-igG_w\hat{A} \right) |\phi\rangle$, where $|\phi\rangle$ is normalized but $|\alpha\rangle$ is not. The expectation of any operator $\hat{M}$ in the final state is, reserved to 1-order with the help of series division:
\begin{align}
\label{derive}
\left\langle \hat{M}\right\rangle _{f}
&=\frac{\langle\alpha|\hat{M}|\alpha\rangle}{\langle\alpha|
\alpha\rangle}, \notag\\
&\simeq\frac{\langle\phi|\left(\hat{M}-igG_w\hat{M}\hat{A} + ig\overline{G}_w\hat{A}\hat{M} \right) |\phi\rangle}{\langle\phi|\left(1-igG_w\hat{A} + ig\overline{G}_w\hat{A} \right) |\phi\rangle},\notag\\
&=\frac{\langle\hat{M}\rangle + \left[ia\langle\hat{A}\hat{M} - \hat{M}\hat{A} \rangle_i +b \langle\hat{A}\hat{M} + \hat{M}\hat{A} \rangle_i \right]g }{1+\left[2b\langle\hat{A}\rangle_i \right] g}, \notag\\
&\simeq \langle\hat{M}\rangle_i + iga\left\langle \hat{A}\hat{M}-\hat{M}\hat{A}\right\rangle _{i}\notag\\&+gb\left\langle \hat{A}\hat{M}+\hat{M}\hat{A}\right\rangle _{i}-2gb\left\langle \hat{M}\right\rangle _{i}\left\langle \hat{A}\right\rangle _{i} ,
\end{align}
which gives Eq.~\eqref{cwv-m}.
Let $\hat{A}=\hat{\sigma}_y$ and put spin components into $M$ in Eq.~\eqref{cwv-m}, we obtain Eq.~\eqref{eqn:complex}.

\begin{figure*}[!htbp] 
\includegraphics[width=17.2cm]{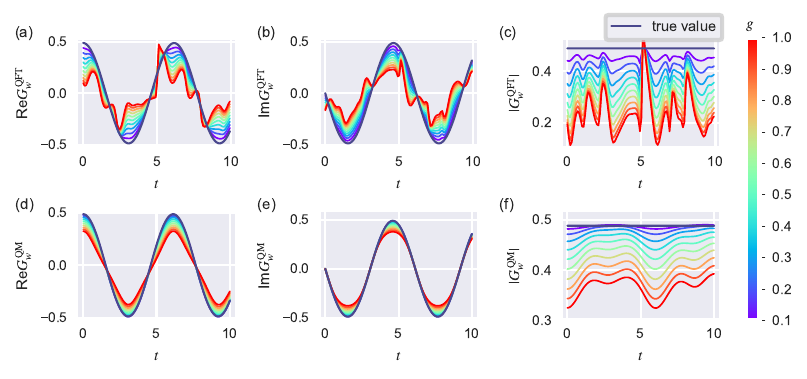}
\caption{\label{S5}  Comparison of one-lattice $\phi^4$ theory with PQHO weak-valued correlation function for different measurement strength $g$ during the continuous time $t\in \left[0,30\right)$. On the upper panels are the simulations of weak measurements in QFT with (a) real part, (b) imaginary part, and (c) module, while on the lower panels are the simulations of weak measurements in QM with (d) real part, (e) imaginary part, and (f) module.}
\end{figure*}

\section{Numerical simulation and detailed trade-off data}\label{NSDTD}

In the numerical simulation, we focus on a specific time slice at $t=5.1$ and consider a set of coupling constants $g\in G=\left\{0.1, 0.2, ..., 1\right\}$ for investigation. The number of copies $N$ and number of repeated experiments $M$ are both integers, with $N$ ranging from 1 to 500000, and $M$ fixed at 10000. Each repeated experiment is assigned a unique random seed ranging from 10 to 10010.

For each experiment, we generate $N$ random numbers following a uniform distribution over the interval $[0, 1)$. These random numbers are used to determine the measured values of $\langle\hat{\sigma}_x\rangle_f$ and $\langle\hat{\sigma}_y\rangle_f$. To calculate $\langle\hat{\sigma}_x\rangle_f$, we assign a value of $-1$ to random numbers that are smaller than the probability $|\langle \sigma_x=-1~|\alpha\rangle|^2 / \langle\alpha|\alpha\rangle$, and $+1$ otherwise. Similarly, for $\langle\hat{\sigma}_y\rangle_f$, we use $|\langle \sigma_y=-1~|\alpha\rangle|^2 / \langle\alpha|\alpha\rangle$ as the threshold to assign $\pm 1$ to the random numbers.

To evaluate the accuracy of the measured results, we focus on the relative deviation from the true value, given by $z_{\text{RV}}=(z_{\text{R}}-z_{\text{TV}}) / |z_{\text{TV}}|$. Utilizing equation Eq.~\eqref{ab}, we obtain $M=10000$ sampling points on the complex plane, denoted as ${ z_{\text{RV}}(m;~g, N),~m\in [0,~M) }$, for each combination of $g\in G$ and $N\in[1,~500000]$.

In the initial step, we study several important statistical quantities to gain a global understanding of the dataset. Specifically, we calculate the mean values and variances for the complex value, module, real part, and imaginary part of ${ z_{\text{RV}}}$. The upper panels of Fig.~\ref{figS2} summarize the results for $g\in G$ and $N\in[1,~5000]$. These panels clearly illustrate the trade-off between $g$ and $N$ as the transition from $N\sim10$ to $N \sim 1000$ becomes apparent.

Additionally, we explore the uncertainty properties, paying particular attention to their dependence on $N$ and $g$. The lower panels of Fig.~\ref{figS2} depict the variance versus the number of copies $1/\delta^2-N$ for the complex value, module, real part, and imaginary part of ${ z_{\text{RV}} }$. Horizontally, the perfect linearity in all four cases verifies our prediction that $\delta \sim 1/\sqrt{N}$. Vertically, the consistent ordering of $g\in G$ for a given $N$ demonstrates that the uncertainty increases with smaller values of $g$.

Finally, we investigate the trade-off between $g$ and $N$ by determining the optimal value of $g\in G$ for different ranges of $N$. Fig.~\ref{Detailed trade-off data} presents the optimal $g\in G$ for each $N\in[1,10000]$. Our criterion for optimality is as follows: for a given $N$, we compare the distances to the true value ${ z_{\text{RV}}(m;~g_i, N)-z_{\text{RV}}(m;~g_j, N),~m\in [0,~M) }$ for any two $g_i,~g_j\in G$. If the total number of cases where $\left[ z_{\text{RV}}(m;~g_i, N)-z_{\text{RV}}(m;~g_j, N)<0 \right]$ is greater than half of the total sampling $M=10000$, we consider $g_i$ to be dominant. The optimal value is the one that dominates over all other $g\in G$. In the case of a draw, we choose the one with the smallest value of $g$.

All codes and data of simulation are available at \href{https://github.com/GnefnAuy/GF-WV}{GitHub}.

\section{Lattice simulation of $\phi^4$ field}\label{LSPF}

We carry out the lattice simulation of scalar $\phi^4$ quantum field in a $(1+1)-d$ discretized  Minkowski space-time, with periodic boundary conditions for both space and time. The space-time lattices are taken as integers given by: $x^\mu = (t,x) = -l, -(l-1), \ldots, l-1, l$ with the lattice spacing $a=1$. We set the initial configuration of $\phi^4$ as:
\begin{align}
\phi_0(x) =& \sqrt{L}\sum_p\frac{\Delta p}{2\pi} \frac{1}{\sqrt{2E_{p}}} \left(  \hat{a}_{p}e^{-ip x} +\hat{a}^{\dagger}_{p}e^{ip x}  \right),\\
\pi_0(x) =& \sqrt{L}\sum_p \frac{-i\Delta p}{2\pi}\sqrt{ \frac{E_{p}}{2}} \left(  \hat{a}_{p}e^{-ip x} -\hat{a}^{\dagger}_{p}e^{ip x}  \right).
\end{align}
where the energy-momentum $E_p={(m^2+p^2)}^{1/2}$, $p = ({2\pi}/{L})x$, the commutation relations $\left[\phi_0(x),\pi_0(y)\right]=i\delta_{xy}$, $\left[\hat{a}_{p},\hat{a}^{\dagger}_{p'} \right]=\delta_{pp'}$, and $L=2l+1$. The field operator of $\phi^4$ could be obtained by switching on time evolution: 
\begin{align}
\phi(t,x) = e^{i\hat{H}t}\phi(0,x)e^{-i\hat{H}t},
\end{align}
where $\hat{H} =1/2 \sum_{x}\Delta x \left[ {\pi_0}^2 + (\nabla \phi_0)^2 +m^2\phi_0^2 + \lambda / {4!} \phi_0^4 \right]$. 

In our numerical simulation, we consider $L=5$, with a truncated Hilbert subspace $N_{\text{cutoff}}=5$ for each individual lattice. The total dimension of the Fock space is $N=5^5=3125$. The ground state $|\Omega\rangle$ is obtained by numerically solving the eigenequations of Hamiltonian. We perform the polar decomposition of the excitation operator $\hat{G}=\mathcal{T} \phi(t,0) \phi(0,0)$ with the singular value decomposition (SVD) method: 
\begin{align}
\hat{A}=\hat{u}\hat{s}\hat{v}=(\hat{u}\hat{v} )(\hat{v}^\dagger sv )=\hat{U}\hat{R}.
\end{align} 
The variation of detection probability in Eq.~\eqref{WMRw} is simulated by the difference between the perturbed and the unperturbed $\delta \text{p} = \text{p}' - \text{p} = |\langle \Theta|\otimes \langle F_a|~e^{-ig\hat{R}\otimes\hat{\sigma}_x}~|\Omega\rangle\otimes| I_a\rangle|^2- |\langle \Theta|\otimes \langle F_a|\Omega\rangle\otimes| I_a\rangle|^2$. Our simulation result of $R_w$ is shown in Fig.~\ref{fig:qft}. Similar to its quantum counterpart, the accuracy of QFT weak measurement increases with small coupling $g$. All codes and data are available at \href{https://github.com/GnefnAuy/GF-WV}{GitHub}.

In order to explore the connection between weak measurements in quantum field theory (QFT) and quantum mechanics (QM), we consider the one-lattice limit where the behavior of the $\phi^4$ field theory resembles that of the PQHO. By numerically solving the eigenequations of the Hamiltonians, we obtain the ground states for both systems. The results, shown in Fig.~\ref{S5}, reveal a good agreement in terms of the true values and trends for different coupling strengths between weak measurements in QFT and QM.
However, it is worth noting that weak measurements in QFT, based on the quantum variation principle, exhibit less accuracy compared to weak measurements in QM under the same conditions. This discrepancy arises due to the approximation $\delta \text{p} = \text{p}' - \text{p}$, which introduces uncertainty in variation $\delta \text{p}$. In the QFT context, this approximation may not hold as well, leading to less precise results in weak measurements compared to QM.
Further investigation and analysis are necessary to understand the specific factors contributing to the differences between weak measurements in QFT and QM. By examining the limitations and approximations involved, we can gain insights into the nature of weak measurements in both frameworks.


\begin{thebibliography}{99}

\bibitem{GFJS} J. Schwinger, On the Green's functions of quantized fields. I, \href{https://www.pnas.org/doi/10.1073/pnas.37.7.452}{Proc. Natl. Acad. Sci. U.S.A. \textbf{37}, 452 (1951).}~J. Schwinger, On the Green's functions of quantized fields. II, \href{https:// www.pnas.org/doi/abs/10.1073/pnas.37.7.455}{Proc. Natl. Acad. Sci. U.S.A. \textbf{37}, 455 (1951).}

\bibitem{QSM} L. P. Kadanoff, Quantum statistical mechanics, CRC Press (2018).

\bibitem{JHEPOTOC} K. Hashimoto, K. Murata, and R. Yoshii, Out-of-time-order correlators in quantum mechanics, \href{https://doi.org/10.1007/JHEP10(2017)138}{J. High Energ. Phys. \textbf{10}, 138 (2017).}

\bibitem{JEOTOC} N. Y. Halpern, Jarzynski-like equality for the out-of-time-ordered correlator, \href{https://journals.aps.org/pra/abstract/10.1103/PhysRevA.95.012120}{Phys. Rev. A \textbf{95}, 012120 (2017).}

\bibitem{scrambling} T. Xu, T. Scaffidi, and X. Cao, Does Scrambling Equal Chaos?, \href{https://journals.aps.org/prl/abstract/10.1103/PhysRevLett.124.140602}{Phys. Rev. Lett. \textbf{124}, 140602 (2020).}

\bibitem{OTOCtopo} Q. Bin, L.-L. Wan, F. Nori, Y. Wu, and X.-Y. L\"u, Out-of-time-order correlation as a witness for topological phase transitions, \href{https://journals.aps.org/prb/abstract/10.1103/PhysRevB.107.L020202}{Phys. Rev. B \textbf{107}, L020202 (2023).}

\bibitem{AAV} Y. Aharonov, D. Z. Albert, and L. Vaidman, How the Result of a Measurement of a Component of the Spin of a
Spin-1/2 Particle Can Turn Out to be 100, \href{https://journals.aps.org/prl/abstract/10.1103/PhysRevLett.60.1351}{Phys. Rev. Lett. \textbf{60}, 1351 (1988).}

\bibitem{bomb} A. C. Elitzur and L. Vaidman, Quantum mechanical interaction-free measurements, \href{https://link.springer.com/article/10.1007/BF00736012}{Foundations of Physics \textbf{23}, 987 (1993).}

\bibitem{cheshire} Y. Aharnov, S. Popescu, D. Rohrlich, and P. Skrzypczyk, Quantum cheshire cats, \href{https://iopscience.iop.org/article/10.1088/1367-2630/15/11/113015}{New J. Phys. \textbf{15}, 113015 (2013).}

\bibitem{dondeestabas} A. Danan, D. Farfurnik, S. Bar-Ad and L. Vaidman, Asking Photons Where They Have Been, \href{https://journals.aps.org/prl/abstract/10.1103/PhysRevLett.111.240402}{Phys. Rev. Lett. \textbf{111}, 240402 (2013)}

\bibitem{foundation} Y. Aharonov, E. Cohen, and A. C. Elitzur, Foundations and applications of weak quantum measurements, \href{https://journals.aps.org/pra/abstract/10.1103/PhysRevA.89.052105}{Phys. Rev. A \textbf{89}, 052105 (2014).}

\bibitem{geometric} E. Sj\"oqvist, Geometric phase in weak measurements, \href{https://www.sciencedirect.com/science/article/abs/pii/S0375960106009169}{Phys. Lett. A \textbf{359}, 187 (2006).}

\bibitem{eraser} M. Cormann, M. Remy, B. Kolaric, and Y. Caudano, Revealing geometric phases in modular and weak values with a quantum eraser, \href{https://journals.aps.org/pra/abstract/10.1103/PhysRevA.93.042124}{Phys. Rev. A \textbf{93}, 042124 (2016).}

\bibitem{discord} L. Li, Q.-W. Wang, S.-Q. Shen, and M. Li, Geometric measure of quantum discord with weak measurements, \href{https://link.springer.com/article/10.1007/s11128-015-1184-9}{Quantum Information Processing \textbf{15}, 291 (2016).}

\bibitem{tomo} S. Wu, State tomography via weak measurements, \href{https://www.nature.com/articles/srep01193}{Sci. Rep. \textbf{3}, 1193 (2013).}

\bibitem{tomosc} L. Qin, L. Xu, W. Feng, and X.-Q. Li, Qubit state tomography in a superconducting circuit via weak measurements, \href{https://iopscience.iop.org/article/10.1088/1367-2630/aa646e}{New J. Phys. \textbf{19}, 033036 (2017).}

\bibitem{steering} R. Uola, A. C. S. Costa, H. C. Nguyen, and O. G\"uhne, Quantum steering, \href{https://journals.aps.org/rmp/abstract/10.1103/RevModPhys.92.015001}{Rev. Mod. Phys. \textbf{92}, 015001 (2020).}

\bibitem{gefen} S. Roy, J. T. Chalker, I. V. Gornyi, and Y. Gefen, Measurement-induced steering of quantum systems, \href{https://journals.aps.org/prresearch/abstract/10.1103/PhysRevResearch.2.033347}{Phys. Rev. Research \textbf{2}, 033347 (2020).}

\bibitem{sqc} Y. Ding, Y. Pan, and X. Chen, Superoscillating Quantum Control Induced By Sequential Selections, \href{https://arxiv.org/abs/2305.04303}{arXiv:2305.04303}

\bibitem{w2s} Y. Pan, J. Zhang, E. Cohen, C.-W. Wu, P.-X. Chen, and N. Davidson, Weak-to-strong transition of quantum
measurement in a trapped-ion system, \href{https://www.nature.com/articles/s41567-020-0973-y}{Nat. Phys. \textbf{16}, 1206 (2020).}

\bibitem{topological} V. Gebhart, K. Snizhko, T. Wellens, A. Buchleitner, A. Romito, and Y. Gefen, Topological transition in measurement-induced geometric phases, \href{https://www.pnas.org/doi/10.1073/pnas.1911620117}{Proc. Natl. Acad. Sci. U. S. A. \textbf{117}, 5706 (2020).}

\bibitem{GL} M. Gell-Mann and F. Low, Bound states in quantum field theory, \href{https://journals.aps.org/pr/abstract/10.1103/PhysRev.84.350}{Phys. Rev. \textbf{84}, 350 (1951).}

\bibitem{tsvf} Y. Aharonov, P. G. Bergmann, and J. L. Lebowitz, Time Symmetry in the Quantum Process of Measurement, \href{https://journals.aps.org/pr/abstract/10.1103/PhysRev.134.B1410}{Phys. Rev. \textbf{134}, B1410 (1964).}

\bibitem{tsvfreview} Y. Aharonov and L. Vaidman, The two-state vector formalism: an updated review, \href{https://arxiv.org/pdf/quant-ph/0105101.pdf}{Time in quantum mechanics, Springer (2018).} Edited by J. G. Muga, R. S. Mayato and I. Egusquiza.

\bibitem{sagawa} T. Sagawa, Weak Value and Correlation Function, \href{https://arxiv.org/pdf/0901.4212.pdf}{arXiv: 0901.4212.}~We notice this preprint after carrying out most of the work, which to our best knowledge, firstly pointed out the relation between weak value and correlation function. However, it aims at showing a new interpretation of the weak value, while we focus on the weak-valued correlation function.

\bibitem{jozsa} R. Jozsa, Complex weak values in quantum measurement, \href{https://journals.aps.org/pra/abstract/10.1103/PhysRevA.76.044103}{Phys. Rev. A \textbf{76}, 044103 (2007).}

\bibitem{dressel} J. Dressel, K. Y. Bliokh, and F. Nori, Classical Field Approach to Quantum Weak Measurements, \href{https://journals.aps.org/prl/abstract/10.1103/PhysRevLett.112.110407}{Phys. Rev. Lett. \textbf{112}, 110407 (2014).}

\bibitem{korotkov2001} A. N. Korotkov, Selective quantum evolution of a qubit state due to continuous measurement, \href{https://journals.aps.org/prb/abstract/10.1103/PhysRevB.63.115403}{Phys. Rev. B \textbf{63}, 115403 (2001).}

\bibitem{jacobs2006} K. Jacobs and D. A. Steck, A straightforward introduction to continuous quantum measurement,  \href{https://www.tandfonline.com/doi/abs/10.1080/00107510601101934}{Contemporary Physics \textbf{47}, 279 (2006).}

\bibitem{pati} A. K. Pati, U. Singh, and U. Sinha, Measuring non-Hermitian operators via weak values, \href{https://journals.aps.org/pra/abstract/10.1103/PhysRevA.92.052120}{Phys. Rev. A \textbf{92}, 052120 (2015).}

\bibitem{QAP} K. A. Milton, Schwinger's Quantum Action Principle, \href{https://arxiv.org/pdf/1503.08091.pdf;}{Springer (2015).}

\bibitem{nonlocal} A. Brodutch and L. Vaidman, Measurements of non local weak values, \href{https://iopscience.iop.org/article/10.1088/1742-6596/174/1/012004}{J. Phys.: Conf. Ser. \textbf{174}, 012004 (2009).}

\bibitem{jwv} K. J. Resch and A. M. Steinberg, Extracting Joint Weak Values with Local, Single-Particle Measurements, \href{https://journals.aps.org/prl/abstract/10.1103/PhysRevLett.92.130402}{Phys. Rev. Lett. \textbf{92}, 130402 (2004).}


\bibitem{bernstein} V. Mnih, C. Szepesv\'ari, and J. Y. Audibert, Empirical bernstein stopping, \href{10.1145/1390156.1390241}{Proceedings of the 25th international conference on Machine learning (2008).}

\bibitem{Mollar} H. Kleinert, Particles and Quantum Fields, \href{https://doi.org/10.1142/9915}{World Scientific (2016)}.

\bibitem{rev} D. Gu\'ery-Odelin, A. Ruschhaupt, A. Kiely, E. Torrontegui, S. Mart\'inez-Garaot, and J. G. Muga, 
Shortcuts to adiabaticity: Concepts, methods, and applications, \href{https://link.aps.org/doi/10.1103/RevModPhys.91.045001}{Rev. Mod. Phys. \textbf{91}, 045001 (2019).}

\bibitem{adolfo} C. W. Duncan and A. del Campo, Shortcuts to adiabaticity assisted by counterdiabatic Born-Oppenheimer dynamics, \href{https://iopscience.iop.org/article/10.1088/1367-2630/aad437}{New J. Phys. \textbf{20}, 085003 (2018).}

\bibitem{dirac} L. Lamata, J. Le\'on, T. Sch\"atz, and E. Solano, Dirac Equation and Quantum Relativistic Effects in a Single Trapped Ion, \href{https://journals.aps.org/prl/abstract/10.1103/PhysRevLett.98.253005}{Phys. Rev. Lett \textbf{98}, 253005 (2007).}

\bibitem{adscft} L. Garc\'ia-\'Alvarez, I. L. Egusquiza, L. Lamata, A. del Campo, J. Sonner, and E. Solano, Digital Quantum Simulation of Minimal AdS/CFT, \href{https://journals.aps.org/prl/abstract/10.1103/PhysRevLett.119.040501}{Phys. Rev. Lett. \textbf{119}, 040501 (2017).}

\bibitem{keldysh} R. van Leeuwen, N. E. Dahlen, G. Stefanucci, C.-O. Almbladh, and U. von Barth, Introduction to the Keldysh Formalism, \href{10.1007/3-540-35426-3_3}{Springer-Verlag Berlin Heidelberg (2006).}

\bibitem{noneqprb2003} A. Cresti, R. Farchioni, G. Grosso, and G. P. Parravicini, Keldysh-Green function formalism for current profiles in mesoscopic systems, \href{https://journals.aps.org/prb/abstract/10.1103/PhysRevB.68.075306}{Phys. Rev. B \textbf{68}, 075306 (2003).}

\bibitem{noneqprd2014} A. Czajka and S. Mr\'owczy\`nski, Ghosts in Keldysh-Schwinger formalism, \href{https://journals.aps.org/prd/abstract/10.1103/PhysRevD.89.085035}{Phys. Rev. D \textbf{89}, 085035 (2014).}

\bibitem{noneqprb2016} M. F. Maghrebi and A. V. Gorshkov, Nonequilibrium many-body steady states via Keldysh formalism, \href{https://journals.aps.org/prb/abstract/10.1103/PhysRevB.93.014307}{Phys. Rev. B \textbf{93}, 014307 (2016).} 

\end{thebibliography}
\end{document}